\documentclass[twocolumn,showpacs,preprintnumbers,amsmath,amssymb]{revtex4}

\usepackage{graphicx}
\usepackage{dcolumn}
\usepackage{bm}

\begin{document}

\preprint{}
\input{epsf.tex}

\epsfverbosetrue

\title{Collective Enhancement and Suppression of Excitation Decay in Optical Lattices}
\author{Hashem Zoubi, and Helmut Ritsch}

\affiliation{Institut f\"{u}r Theoretische Physik, Universit\"{a}t Innsbruck, Technikerstrasse 25, A-6020 Innsbruck, Austria}

\date{23 October, 2009}

\begin{abstract}
We calculate radiative lifetimes of collective electronic excitations of atoms in an infinite one dimensional lattice. The translational symmetry along the lattice restricts the photon wave vector component parallel to the lattice to the exciton wave number and thus the possible emission directions. The resulting radiation damping rate and emission pattern of the exciton strongly deviates from independent atom. For some wave numbers and polarizations the excitons superradiantly decay very fast, while other excitons show zero radiation damping rate and form propagating meta-stable excitations. Such states could be directly coupled via tailored evanescent fields from a nearby fiber.
\end{abstract}


\maketitle

Ultracold atoms in optical lattices are now one of the most flourishing directions of experimental and theoretical quantum physics \cite{Dalibard}. Beside fundamental physics issues on entanglement, measurement and decoherence, more and more applications and connections appear as test-systems for a wide range of puzzling effects in solid state physics. The trapped atoms can be considered as artificial crystals, which differ from solid crystals in their precise and easy controllability of occupation number, lattice constant, lattice depth and symmetry. The lattice properties can be controlled through the laser intensity, wave length and polarizations, and through laser cooling the motion can be cooled and confined to the lowest state in each potential well corresponding to the first Bloch band. The atomic dynamics is generally well described by a Bose-Hubbard model \cite{Jaksch}, which includes atom hopping among nearest neighbor sites and the on-site repulsive atom-atom interactions. As central prediction one gets a quantum phase transition from a superfluid into a Mott insulator phase \cite{Greiner,Jaksch} with a fixed number of atoms per site for a deep lattice. Other condensed matter phenomena appearing in more complex situations can be studied in generalized optical lattice setups as well \cite{Lewenstein}.

Optical properties of crystal solids are a well studied subject. One of the phenomena that strongly characterize the optical properties of solids is the formation of collective electronic excitations (excitons). In molecular crystals a local electronic excitation delocalizes among the lattice molecules due to electrostatic interactions. The corresponding delocalized eigenstates are Frenkel excitons \cite{Davydov}, which are quasi-particles that propagate in the lattice. In previous work we showed that Frenkel like-excitons also exist for ultracold atom lattices in the Mott insulator phase \cite{ZoubiA,Antezza}. Enclosing the light field within a cavity, such excitons and photons with the same wave vectors coherently couple to form cavity polaritons \cite{ZoubiA,ZoubiC}.

In the present work we study the radiative properties of such excitons, which are formed due to resonant dipole-dipole interactions. Each exciton is represented by a wave that propagates in the lattice with a given wave number \cite{ZoubiA} and coupled to the free space radiation field modes into which it can decay. For the moment we restrict ourselves to an infinite one dimensional lattice with one atom per site. This allows one to directly calculate the far field radiation pattern and the resulting effective damping rate of the exciton explicitly. As a result of the lattice symmetry an exciton with a fixed wave number can emit a free photon with the same wave number component along the lattice axis. In particular we investigate, how the damping rate, which significantly differs from the independent atom result, changes with the wave number and polarization direction.

\ 

We consider a one-dimensional string of two-level atoms with an electronic transition energy $E_a=\hbar\omega_a$ with lattice constant $a$ and one atom per site (see figure (1)). Energy is transferred among atoms at different sites by resonant dipole-dipole interactions \cite{ZoubiA}, so that the electronic excitation Hamiltonian contains two terms, the on-site excitation and the energy transfer, it reads
\begin{equation}
H_{ex}=\sum_n\hbar\omega_a\ B_n^{\dagger}B_n+\sum_{nm}\hbar J_{nm}(\theta)\ B_n^{\dagger}B_m,
\end{equation}
where $B_n^{\dagger}$ and $B_n$ are the electronic excitation creation and annihilation operators at site $n$, respectively, which are assumed to be boson operators at low excitation density with the commutation relation $[B_n,B_m^{\dagger}]=\delta_{nm}$. The Hamiltonian can be easily diagonalized, in exploiting the lattice symmetry, by using the transformation
\begin{equation} \label{TRANS}
B_n=\frac{1}{\sqrt{N}}\sum_ke^{ikz_n}B_k,
\end{equation}
where $N$ is the number of sites, and $z_n=an$ is the position of site $n$. We get the collective electronic excitation Hamiltonian
\begin{equation}
H_{ex}=\sum_k\hbar\omega_{ex}(k,\theta)\ B_k^{\dagger}B_k,
\end{equation}
with the dispersion $\omega_{ex}(k,\theta)=\omega_a+\sum_LJ(L,\theta)e^{ikL}$, and $L=(n-m)a$. Restricting the energy transfer to only nearest neighbor sites, we explicitly obtain $\omega_{ex}(k,\theta)=\omega_a+2J(\theta)\ \cos(ka)$, where we used $J(\theta)=J(a,\theta)$, and for the dipole-dipole interaction we have $\hbar J({\theta})=\frac{\mu^2}{4\pi \epsilon_0a^3}\left(1-3\cos^2\theta\right)$, here $\mu$ is the electric transition dipole, and $\theta$ is the angle between the transition dipole and the lattice axis, as seen in figure (1). Including corrections from more distant neighboring atoms will change this spectrum and the particular form of the excitons, which cannot be calculated explicitly any more. Nevertheless, the results and conclusions of the following life time calculations are generally only weakly influenced by these corrections as the wave nature of the exciton is mainly determined by symmetry properties of the lattice. In using periodic boundary condition, the wave number $k$ takes the values $k=\frac{2\pi}{Na}p$, with $p=0,\pm 1,\pm 2,\cdots,\pm N/2$. It is clear that in place of the discrete atomic transition we get an energy band of band width $4J(\theta)$.

Due to the energy transfer, an quasi-stationary electronic excitation is delocalized in the lattice and best represented by a wave that propagates to the left or the right direction with wave number $k$, which is a good quantum number. Such collective electronic excitations are called ``excitons''. While some excitons directly couple to incoming plane waves, for other excitons  with a fixed wave number $k$ and a prescribed transition dipole direction $\theta$, such a straightforward excitation is not possible. As has been implemented recently, one way to overcome this issue is the use of a thin pulled fiber along the lattice \cite{Vetsch}. The atomic lattice is generated parallel to the fiber at given distance, where the fiber photons and the corresponding lattice excitons are coupled. Hence, by sending photons with prescribed wave number and polarization through the fiber we can excite an exciton in the lattice with the same wave number and polarization through the evanescent wave coupling \cite{ZoubiB}.

\begin{figure}[h!]
\centerline{\epsfxsize=6cm \epsfbox{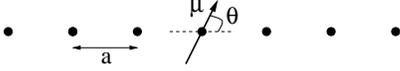}}
\caption{A one-dimensional lattice of lattice constant $a$, and with one atom per site. The transition dipole $\mu$ is plotted which makes an angle $\theta$ with the lattice axis.}
\end{figure}

Here we calculate the electric field and intensity of the radiation emitted by a 1D lattice exciton into free space. The free-space radiation field Hamiltonian is given by
\begin{equation}
H_{rad}=\sum_{{\bf q}\lambda}E_{ph}(q)\ b_{{\bf q}\lambda}^{\dagger}b_{{\bf q}\lambda},
\end{equation}
where $b_{{\bf q}\lambda}^{\dagger}$ and $b_{{\bf q}\lambda}$ are the creation and annihilation operators of a photon with wave vector ${\bf q}$ and polarization $\lambda$, respectively. The photon energy is $E_{ph}(q)=\hbar\omega_r(q)=\hbar cq$. The electric field operator is
\begin{equation} \label{EField}
\hat{\bf E}_{rad}({\bf r})=i\sum_{{\bf q}\lambda}\sqrt{\frac{\hbar cq}{2\epsilon_0 V}}\left\{b_{{\bf q}\lambda}\ {\bf e}_{{\bf q}\lambda}e^{i{\bf q}\cdot{\bf r}}-b_{{\bf q}\lambda}^{\dagger}\ {\bf e}_{{\bf q}\lambda}^{\ast}e^{-i{\bf q}\cdot{\bf r}}\right\},
\end{equation}
where ${\bf e}_{{\bf q}\lambda}$ is the photon polarization unit vector, and $V$ is the normalization volume.

The light-matter coupling is given by the electric dipole interaction $H_I=-\hat{\mu}\cdot\hat{\bf E}_{rad}$, where the atomic transition dipole operator is defined by $\hat{\mu}=\vec{\mu}\sum_n\left(B_n+B_n^{\dagger}\right)$. We separate the photon wave vector into two components, one parallel to the lattice axis and the other orthogonal to it, namely we use ${\bf q}={\bf q}'+q_z{\bf z}$, with $q^2=q^{\prime 2}+q_z^2$. The atomic site positions are given by ${\bf r}_n=z_n{\bf z}$. We represent the electronic excitations by excitons, in using the inverse transformation of equation (\ref{TRANS}), and by using the lattice symmetry property $\frac{1}{N}\sum_ne^{i(q_z-k)z_n}=\delta_{q_zk}$. The coupling between the excitons and the free radiation field in the rotating wave approximation and with linear polarizations, is given by
\begin{equation}
H_I=\sum_{{\bf q'}k\lambda}i\hbar g_{{\bf q}\lambda}\left\{b_{{\bf q'}k,\lambda}B_k^{\dagger}-b_{{\bf q'}k,\lambda}^{\dagger}B_k\right\},
\end{equation}
where now ${\bf q}=({\bf q'},k)$, with the coupling parameter
\begin{equation}
\hbar g_{{\bf q}\lambda}=-\sqrt{\frac{\hbar cqN}{2\epsilon_0 V}}\left(\vec{\mu}\cdot{\bf e}_{{\bf q}\lambda}\right).
\end{equation}
It is seen that the coupling is only between excitons and photons with the same $z$ component wave vectors. The translational symmetry along the lattice axis results in a conservation of momentum along the axis. Hence an exciton with wave number $k$ couples only to photons with the same $k$ parallel to the lattice. Such a result strongly affect both the radiation emitted by an exciton with a fixed $k$, and the damping rate of such exciton, as we show in the following. The effect of similar retarded interaction on the exciton spectrum at low dimensional molecular crystals was studied by Agranovich et. al. \cite{Agranovich}.

Now we calculate the electric field and the intensity of the light emitted from one dimensional optical lattice of cold atoms. We start from the radiation field operator equation of motion, solving it formally and substituting the source part back in the electric field operator of equation (\ref{EField}), then we obtain for the positive part of the electric field, for a fixed $k$ and $\theta$,
\begin{eqnarray}
&&\hat{\bf E}^+_{rad}({\bf r},t)=i\sum_{{\bf q'}\lambda}\frac{\omega_r(q)\sqrt{N}}{2\epsilon_0 V}\ e^{i\left[{\bf q'}\cdot\vec{\rho}+kz-\omega_e(k)t\right]}\times \nonumber \\
&&{\bf e}_{{\bf q}\lambda}\left(\vec{\mu}\cdot{\bf e}_{{\bf q}\lambda}\right)\ \int_0^{t}dt'\ \tilde{B}_k(t')\ e^{i\left[\omega_e(k)-\omega_r(q)\right](t-t')},
\end{eqnarray}
where the exciton operator in a rotating frame is defined by $B_k(t)=\tilde{B}_k(t)\ e^{-i\omega_e(k)t}$, and the observation point is at ${\bf r}=\vec{\rho}+z{\bf z}$. As the wave vector component parallel to the lattice is fixed, we have only to sum over wave vectors in the plane normal to the lattice direction. The sum over ${\bf q'}$ casts into the integral
\begin{equation}\label{intsum}
\sum_{{\bf q}'}\rightarrow \frac{S}{4\pi^2}\int d^2q'=\frac{S}{4\pi^2}\int_{-\pi}^{+\pi}d\phi\int_0^{\infty}q' dq',
\end{equation}
where $V=SL$, and $S$ is the normalization plane, with the lattice length $L=Na$. We use also the summation over the photon polarization $\sum_{\lambda}{\bf e}_{{\bf q}\lambda}{\bf e}_{{\bf q}\lambda}=1-\frac{{\bf q}{\bf q}}{q^2}$. For the different vectors we choose: the wave vector is ${\bf q}=\left(q'\cos\phi,q'\sin\phi,k\right)$, the transition dipole is $\vec{\mu}=\left(\mu\sin\theta,0,\mu\cos\theta\right)$, and the observation point is chosen to be at ${\bf r}=\left(\rho,0,z\right)$, as seen in figure (2).

After doing the angular integral we keep only the far field terms proportional to $1/\sqrt{\rho}$. As solvable example we consider long wave length (small wave number) excitons with $ka\ll 1$. Then we apply the Weisskopf-Wigner approximation, which is in the spirit of the Markov approximation \cite{Wolf}. The retarded electric field is given by
\begin{equation}
\hat{\bf E}^+_z({\bf r},t)=(1+i)\frac{\mu\omega_e(k)^{3/2}\cos\theta}{4\pi\epsilon_0ac^{3/2}}\sqrt{\frac{\pi}{N\rho}}\ B_k(t-\rho/c)\ e^{ikz},
\end{equation}
and the intensity operator is
\begin{equation}
\hat{I}=\hat{\bf E}^-_z\hat{\bf E}^+_z=\left(\frac{\mu\cos\theta}{4\epsilon_0a}\right)^2\frac{2\omega_e(k)^3}{N\pi\rho c^3}\ B_k^{\dagger}(t-\rho/c)B_k(t-\rho/c).
\end{equation}
The expectation values of the exciton operator is
\begin{equation}
\langle B_k(t-\rho/c)\rangle=\langle B_k(0)\rangle\ e^{-i\omega_e(k)(t-\rho/c)}\ e^{-\Gamma_k(t-\rho/c)/2},
\end{equation}
where $\Gamma_k$ is the damping rate of an exciton with wave number $k$. The expectation value of the intensity is
\begin{equation}
\langle\hat{I}\rangle=\left(\frac{\mu\cos\theta}{4\epsilon_0a}\right)^2\frac{2\omega_e(k)^3}{N\pi\rho c^3}\ \langle B_k^{\dagger}(0)B_k(0)\rangle\ e^{-\Gamma_k(t-\rho/c)}.
\end{equation}
Our next task should be to calculate the exciton damping rate, $\Gamma_k$, into free space, and to compare the result with the single atom damping rate in order to emphasize the cooperative effect among the atoms on the damping rate.

\begin{figure}[h!]
\centerline{\epsfxsize=3.5cm \epsfbox{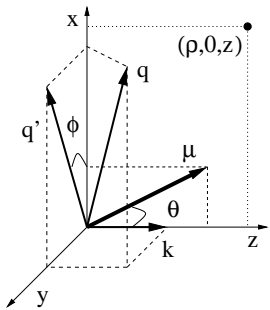}}
\caption{The lattice is in the $z$ direction, where the transition dipole $\mu$ is in the $(x-z)$ plane and makes and angle $\theta$ with the lattice direction. The exciton has a wave number $k$ along the lattice. The photon wave vector ${\bf q}$ has a component ${\bf q'}$ normal to the lattice in the $(x-y)$ plane, and a $z$ component $k$ parallel to the lattice which is equal to the exciton wave number. The observation point is at $(\rho,0,z)$.}
\end{figure}

The radiative exciton damping rate with a fixed $k$ can be evaluated using the Fermi Golden Rule, which reads
\begin{equation}
\Gamma_k=\frac{2\pi}{\hbar}\sum_{{\bf q}'\lambda}\left|\langle f|H_I|i\rangle\right|^2\delta\left(E_{ex}(k)-E_{ph}(q)\right).
\end{equation}
The initial state is of a single exciton of wave number $k$ in the lattice and an empty radiation field, that is $|i\rangle=|1_{ex}(k),0_{ph}\rangle$, and the final state is of zero excitons and a single radiation field photon of wave vector ${\bf q}$, that is $|f\rangle=|0_{ex},1_{ph}({\bf q})\rangle$. The matrix element is given by $\langle f|H_I|i\rangle=i\sqrt{\frac{\hbar cqN}{2\epsilon_0 V}}\left(\vec{\mu}\cdot{\bf e}_{{\bf q}\lambda}\right)$. As before the emitted photon has a wave vector component parallel to the lattice equals to the exciton wave number $k$. The summation is over the wave vectors ${\bf q'}$ normal to the lattice direction. The delta function cares for the conservation of energy of the exciton and the emitted photon. The summation over the photon polarizations is obtained by the relation $\sum_{\lambda}\left|\vec{\mu}\cdot{\bf e}_{{\bf q}\lambda}\right|^2=\left|\vec{\mu}\right|^2-\frac{\left|{\bf q}\cdot\vec{\mu}\right|^2}{q^2}$. The sum over ${\bf q}'$ is converted into the integral of equation (\ref{intsum}). As before we assume a transition dipole with a fixed direction of $\vec{\mu}=(\mu\sin\theta,0,\mu\cos\theta)$, and with ${\bf q}\cdot\vec{\mu}=q'\mu\sin\theta\cos\phi+k\mu\cos\theta$. After integration we obtain the result
\begin{equation} \label{Damping}
\Gamma_k=\frac{\mu^2E_{ex}^2(k)}{4\epsilon_0a\hbar^3c^2}\left\{1+\cos^2\theta-\frac{(\hbar ck)^2}{E_{ex}^2(k)}\left(2\cos^2\theta-\sin^2\theta\right)\right\}.
\end{equation}
Note that the free atom damping rate is $\Gamma_{at}=\frac{\mu^2E_a^3}{3\pi\epsilon_0\hbar^4c^3}$.

As expected for long enough wave vectors $k$ and certain angles $\theta$ the damping rate can be much larger than the single atom one, as we get a superradiant enhancement analogous to the Dicke model. However, we see that for some $k$ and $\theta$ the damping rate can be much smaller than the single atom one or even completely vanish. Such metastable states (dark states), which appear for lattices with two atoms per site \cite{ZoubiC}, thus also exist in infinite chains of atoms as treated. Quite generally at $\theta=0^o$  we can calculate a critical wave vector $k_c$, given by $E_{ex}(k_c)=\hbar ck_c$, above which the damping rate is zero and the excitations can no longer decay radiatively. For general polarization such a critical wave vector exists if the following equation has a solution
\begin{equation}
\frac{(\hbar ck_c)^2}{E_{ex}^2(k_c)}=\frac{1+\cos^2\theta}{2\cos^2\theta-\sin^2\theta}.
\end{equation}

\begin{figure}[h!]
\centerline{\epsfxsize=4cm \epsfbox{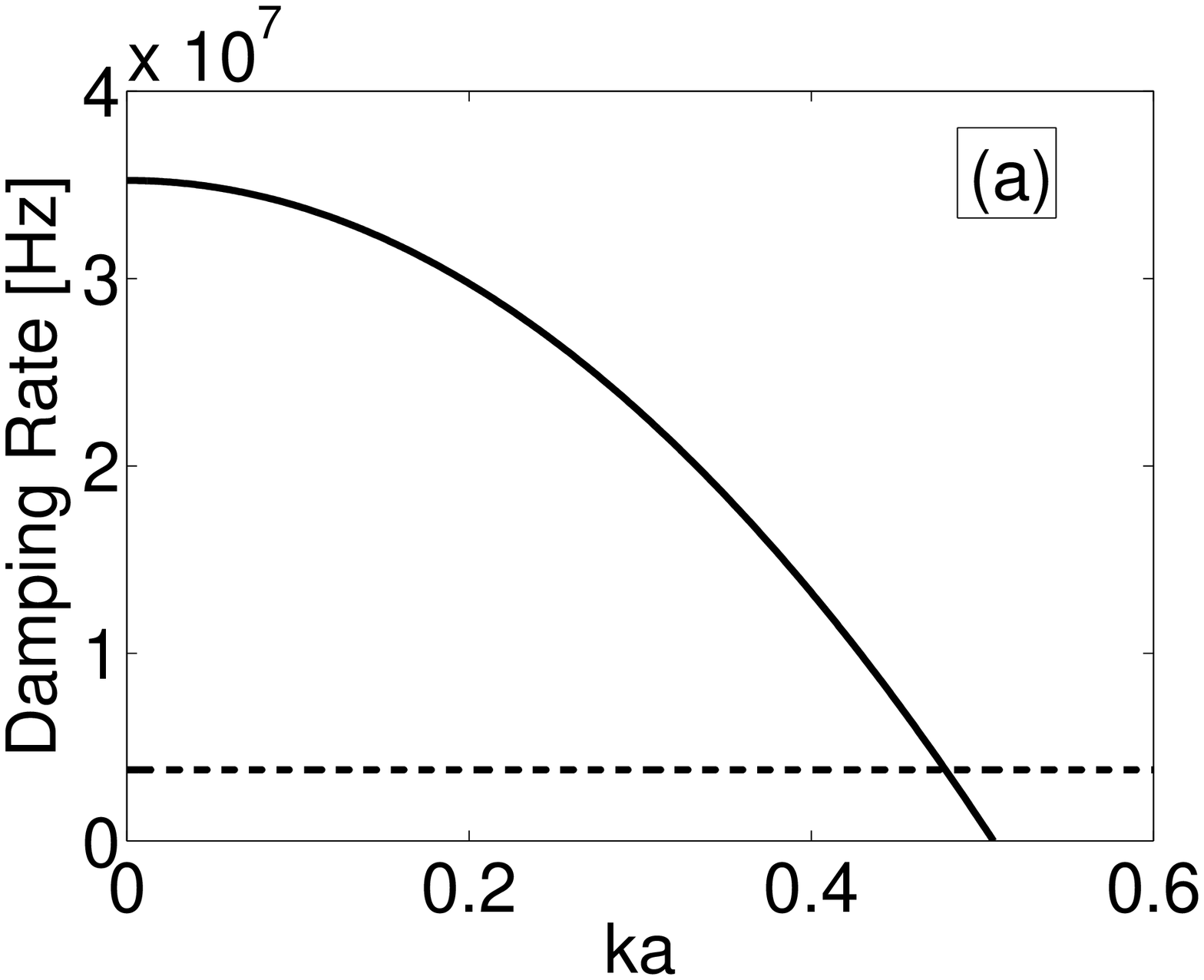}\ \ \ \epsfxsize=4cm \epsfbox{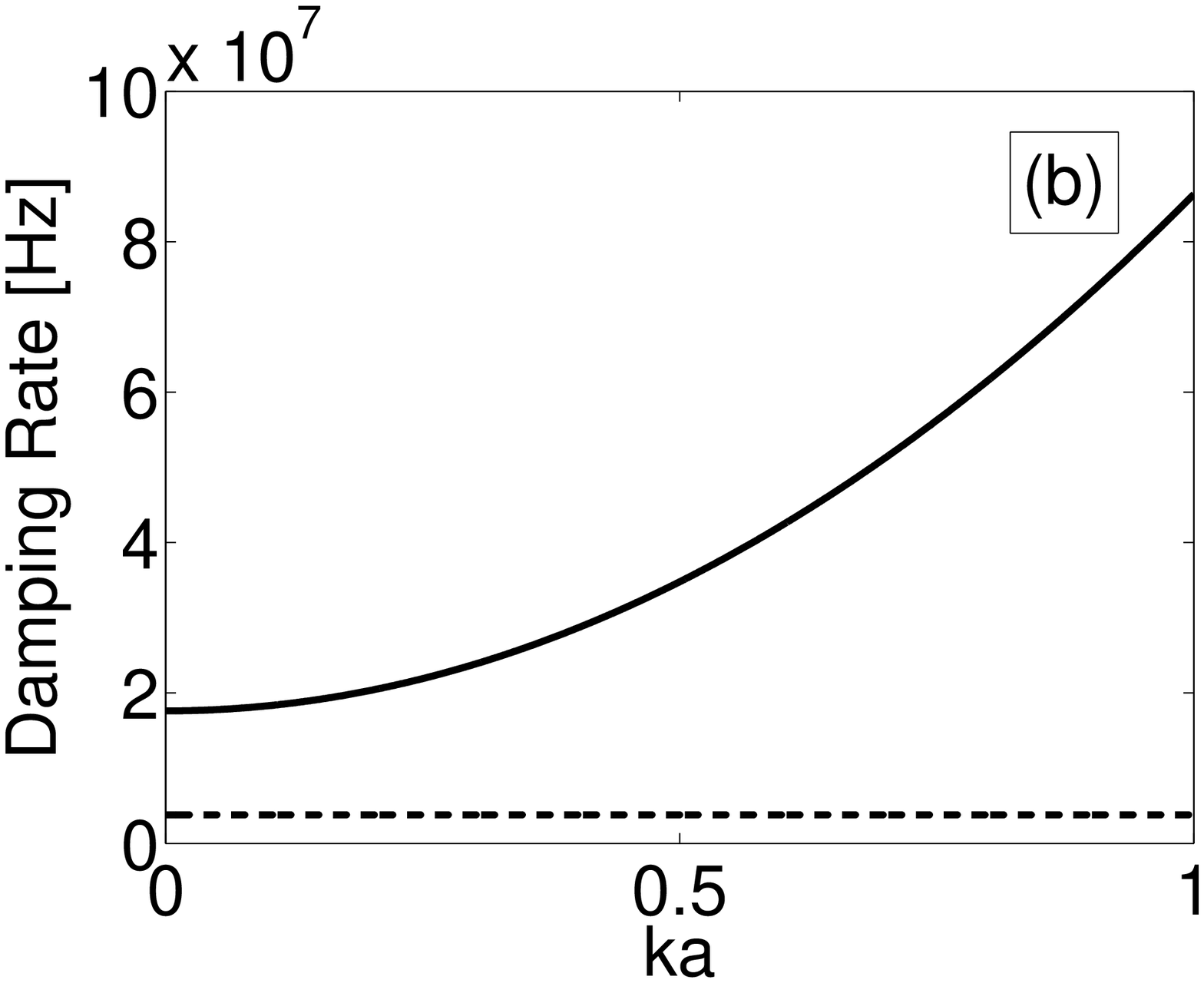}}
\caption{The damping rate vs. $ka$, for (a) $\theta=0$, and (b) $\theta=90$. The dashed-line is for a single atom damping rate.}
\end{figure}

 Let us now exhibit this behavior in several plots of the damping rate. We use the parameters: the transition energy is $E_a=1\ eV$, the lattice constant is $a=1000\ A$, and the transition dipole is $\mu=1\ eA$. In figure (3.a) we plot the damping rate as a function of $ka$ for $\theta=0^o$, the plot includes also the damping rate of a single atom (dashed line). Note that for small wave vectors, $ka\sim 0$ the exciton damping rate is much larger than the free atom damping rate, and these states are superradiant states. With increasing wave vectors the damping rate drops and reaches the single atom rate, and beyond a critical wave vector $k_c$ the damping rate becomes zero and no damping is obtained beyond $k_c$. In figure (3.b) the plot is for $\theta=90^o$, here the damping rates are larger than the single atom one for small wave vectors, $ka\sim 0$, but now the damping rate increases with increasing the wave vector. Namely, the large wave vector states became more superradiant states. Exist an angle, around $\theta=54.7^o$, where the damping rate changes its behavior.

\begin{figure}[h!]
\centerline{\epsfxsize=4cm \epsfbox{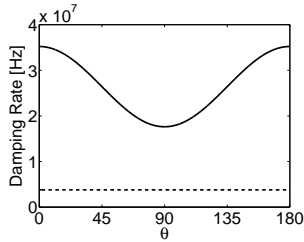}}
\caption{The damping rate vs. $\theta$ (full-line), for $ka=0.01$. The dashed-line is for a single atom damping rate.}
\end{figure}

\begin{figure}[h!]
\centerline{\epsfxsize=4cm \epsfbox{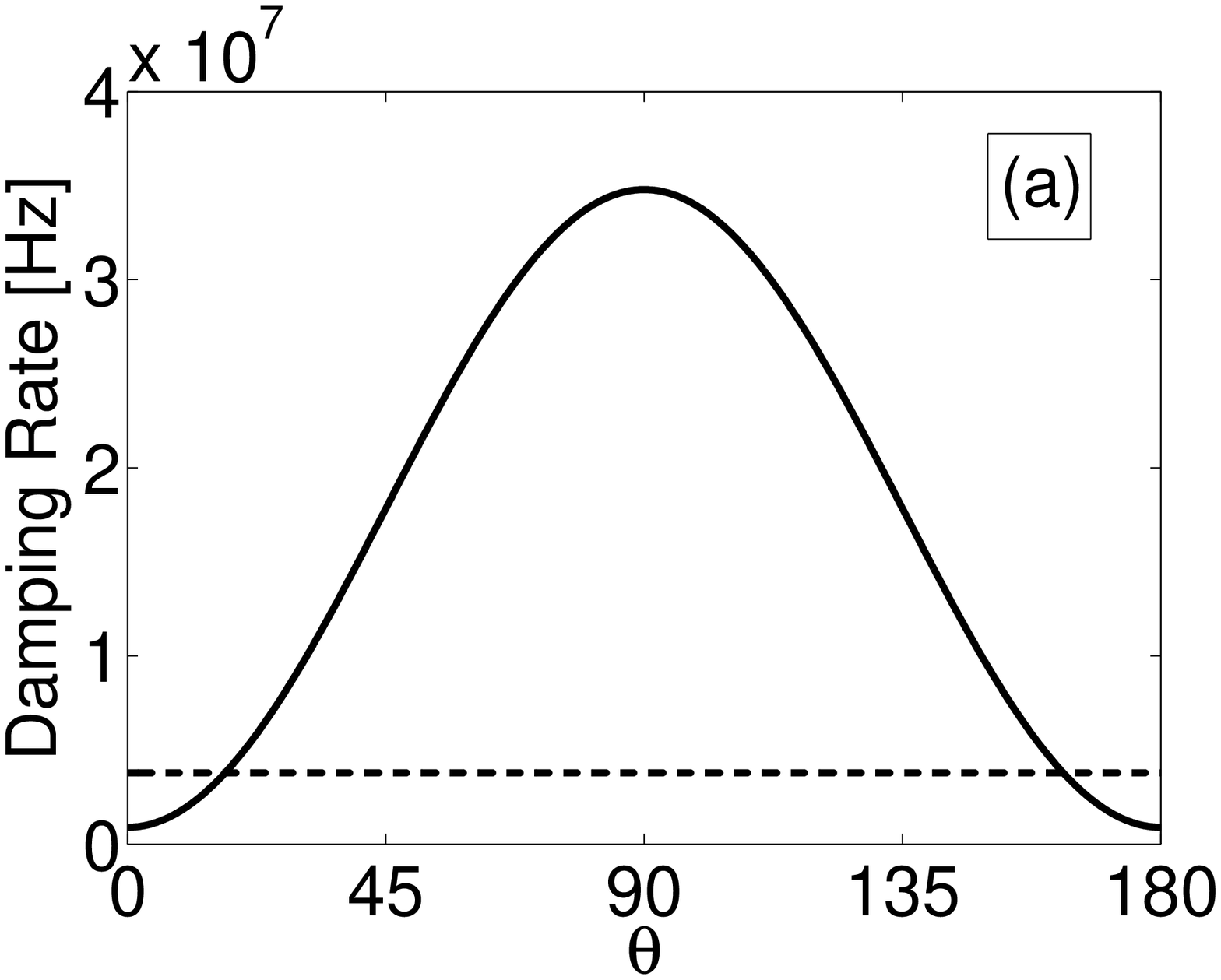}\ \ \ \epsfxsize=4cm \epsfbox{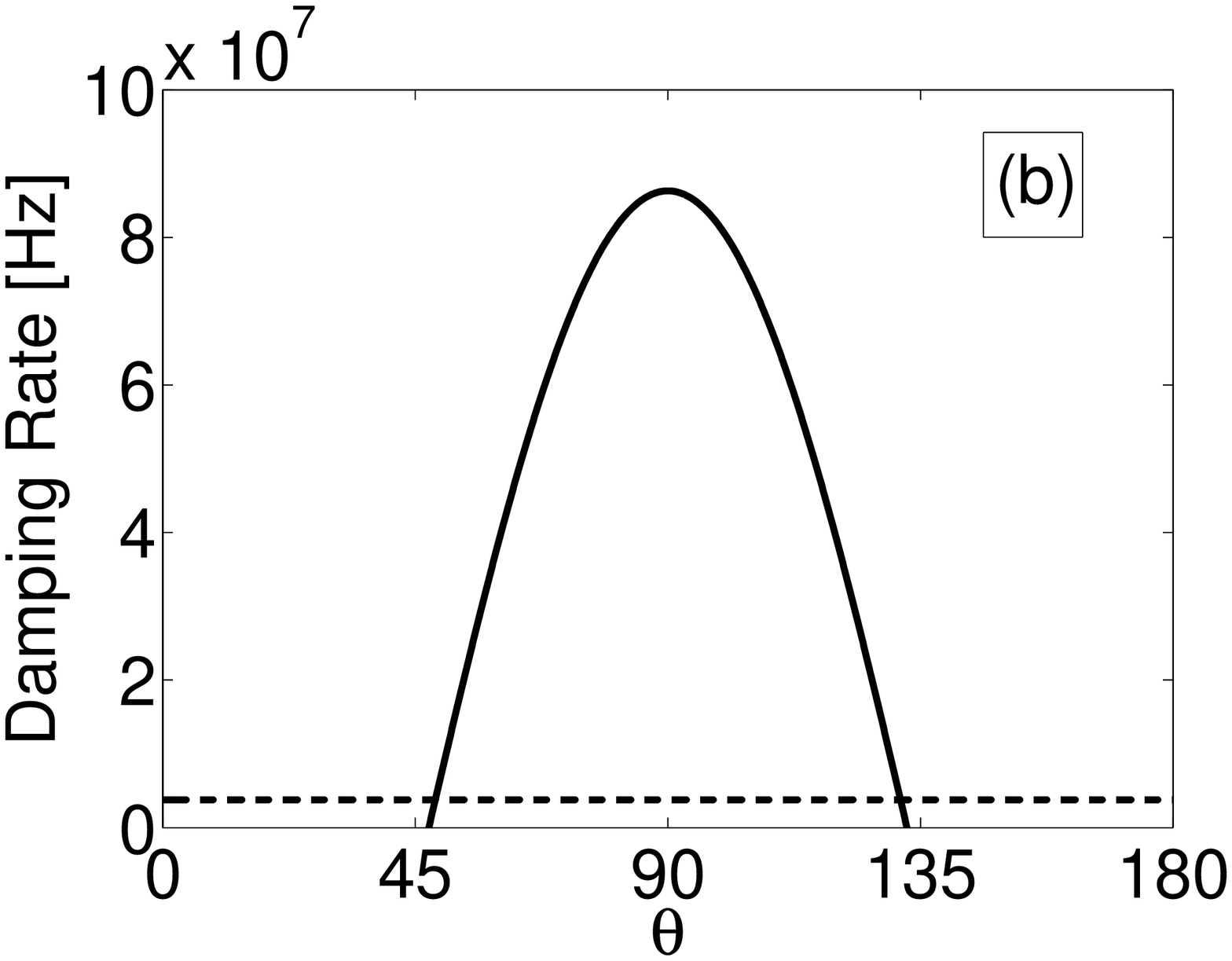}}
\caption{The damping rate vs. $\theta$, for (a) $ka=0.5$, and (b) $ka=1$. The dashed-line is for a single atom damping rate.}
\end{figure}

Next for different values of $ka$ we plot the damping rate as a function of $\theta$. In figure (4) we use $ka=0.01$, here the damping rate is larger than a single atom one for all angles, but around $\theta=90^o$ the damping rate is lower than that of $\theta=0^o$. In figure (5.a) we use $ka=0.5$, now a dramatic change is obtained, where the damping rate is smaller than the single atom one around $\theta\sim 0^o$ and increases to a maximum at $\theta=90^o$, that is much larger than the single atom case and which is strongly superradiant state. For larger $ka$, in figure (5.b) we use $ka=1$, now for angles up to $45^o$ the damping rate is zero, and they are nonzero for $\theta \sim 45^o-135^o$ with a high maximum at $\theta=90^o$. For $ka=\pi$ at the boundary of the Brillouin zone, the non-zero region is for $\theta \sim 54.7^o-125.3^o$, and with higher rate at $\theta=90^o$ than the previous case of figure (5.b).

\ 

We showed that the damping rate and the radiation emitted of an exciton in one dimensional lattice is strongly different from independent atoms. The excitons formed by the energy transfer among the lattice atoms due to dipole-dipole interactions can be characterized by their wave numbers along the lattice. This restricts the spontaneously emitted photons to the same wave vector component parallel to the lattice. The radiative damping rate of such excitons into free space strongly depends on both, the wave number and the polarization direction of the exciton. In close analogy to superfluorescent scattering, some wave numbers and polarizations exhibit a much larger damping rate than an independent atom (superradiant excitons), while for other wave numbers and polarizations the damping rate is much smaller and even can vanish completely, so that the excitons now are metastable. The present results can be adopted to any periodic distribution of optically active materials, e.g. a lattice of quantum dots, one dimensional molecular crystals, periodic semiconductor nanostructures, or a chain of trapped ions, etc. Interestingly such excitons might still be able to couple to and decay into a nearby material structure if it supports the corresponding wave vectors. As a single atom can only bear one excitation, several of these excitons will interact and might open possibilities of nonlinear light interactions at very low intensities.

The work was supported by the Austrian Science Funds (FWF), via the project (P21101 and S40130). We thank A. Rauschenbeutel for communicating his recent experimental results \cite{Vetsch}.


\begin{thebibliography}{}
\bibitem{ZoubiB} More details concerning an atomic lattice parallel to a cavity fiber will be released by us in the near future.
\end{thebibliography}
\end{document}